\documentclass[prb,twocolumn]{revtex4}

\usepackage{graphicx}

\begin{document}

\title{Phase separation in the Hubbard model}
\author{
A.\ Macridin$^{1}$,  M. \  Jarrell$^{1}$, Th.\ Maier$^{2}$}

\address{
$^{1}$University of Cincinnati, Cincinnati, Ohio, 45221, USA \\
$^{2}$Oak Ridge National Laboratory, Oak Ridge, Tennessee, 37831, USA}

\date{\today}

\begin{abstract}
 
Phase separation in the Hubbard model is investigated with the dynamical 
cluster approximation.  We find that it is present in the paramagnetic 
solution  for values of filling smaller than one and at finite 
temperature when a positive next-nearest neighbor hopping is considered. 
The phase separated region is characterized by a mixture of a strongly 
correlated metallic  and   Mott insulating  phases. 
Our results indicate that phase separation is driven by the formation of doped 
regions with strong antiferromagnetic correlations and low kinetic energy.

\end{abstract}

\maketitle
\paragraph*{Introduction}

There is  strong experimental evidence that  high T$_c$ materials are 
susceptible to charge inhomogeneities, such as stripes~\cite{stripe} or 
checkerboard  modulation~\cite{stm}.  This discovery has spurred great 
theoretical interest in phase separation (PS) in models related to the 
cuprates, such as the Hubbard model which is believed to capture the 
low-energy physics of cuprate superconductors.  It was argued by different 
authors that the charge instability displayed as PS in such 
simple models without long-range Coulomb interaction evolves into 
incommensurate charge ordering when the long-range repulsion is 
considered~\cite{PStoCDW}.   In this paper we present results on  
PS in the Hubbard model.   We find that the paramagnetic
asymmetric Hubbard model near half filling phase separates into 
undoped Mott liquid and doped Mott gas phases.  The resulting Mott 
liquid-Mott gas phase diagram bears a strong resemblance to that of a 
classical liquid gas mixture.

Phase separation in the Hubbard and in the closely related t-J model
has been intensively investigated. There is a general consensus that a 
t-J model with a large $J/t$ separates into two phases, an 
undoped antiferromagnet (AF) and a hole rich region. However the results 
for realistic  $J/t < 1$ are controversial. Emery {\em et al.}~\cite{EKL},
Hellberg {\em et al.}~\cite{HellbergManousakis} and Gimm 
{\em et al.}~\cite{u1sboson} report PS for all  values of  
$J/t$.  Others authors such as  Putikka {\em et al.}~\cite{putikka}
and Shih {\em et al.}~\cite{shih} find  no PS for small   
$J/t$. In the Hubbard model with only nearest-neighbor hopping, 
exact diagonalization~\cite{moreo}
and Monte Carlo~\cite{sorella} calculations show no evidence of
PS. These numerical  results are  consistent with the 
analytical results of G. Su's~\cite{su}, who show that there is no phase  
separation in the particle-hole symmetric Hubbard model. However, a 
large-$N$ investigation of this model in the infinite $U$ limit 
shows PS when the next-nearest
neighbor hopping $t'$ is considered~\cite{gehlhoff}. Phase separation in 
the Hubbard model at small doping was also found in a dynamical mean 
field calculation in the antiferromagnetic phase~\cite{pruschke} and with 
variational cluster perturbation theory \cite{arrigoni} in the 
antiferromagnetic and superconducting phases.

Phase separation is believed to be closely related with the antiferromagnetic 
order; a homogeneous  doped system is unstable preferring to separate into 
an undoped antiferromagnetic region  which lowers the exchange energy 
(maximizes the number of antiferromagnetic bonds) and a rich doped phase 
with low kinetic energy. The driving force for PS in a t-J model when 
$J/t$ is large will therefore be  the desire to form undoped antiferromagnetic
regions~\cite{EKL}. However in the Hubbard model 
we did not find  PS for the  values of parameters which are optimal
for antiferromagnetic order in the undoped region.
For instance,  with DCA the maximum Ne\'el temperature 
in the undoped system is obtained for $U \approx 3/4 W$, $W=8t$ 
being the electronic bandwidth, and 
for $t'=0$. The later conditions can be understood by noticing
that  a finite $t'$  introduces an antiferromagnetic exchange between 
the same  sublattice sites, thus frustrating the antiferromagnetism. 
Nevertheless we find PS only for  a $U \ge W$ and a 
finite next-nearest-neighbor hopping $t' \ne 0$.
Moreover, we find  PS  in the paramagnetic solution  which shows that short 
range antiferromagnetic correlations are sufficient for the PS to take place. 
Presumably the PS is driven by the formation of weakly doped regions 
with strong antiferromagnetic correlations and low kinetic energy.
The main culprit for the low value of the kinetic energy is the parameter $t'$ 
with the right sign.

\paragraph*{Formalism}

We use the Dynamical Cluster Approximation (DCA)\cite{hettler:dca,maier:rev} 
to explore the possibility of PS in the 2D Hubbard model, 
with 
\begin{equation} 
\label{eq:HM}
H=H_{kin}+H_{pot}
\end{equation}
\noindent where
\begin{eqnarray} 
\label{eq:hams}
H_{kin}&=&-t \sum_{\langle ij\rangle, \sigma} c^\dagger_{i\sigma}c^{\phantom\dagger}_{j\sigma}
-t' \sum_{\langle\langle il\rangle \rangle, \sigma} c^\dagger_{i\sigma}
c^{\phantom\dagger}_{l\sigma}\\
H_{pot}&=& U\sum_i n_{i\uparrow} n_{i\downarrow}~.
\end{eqnarray}
\noindent Here $c^{(\dagger)}_{i\sigma}$ (creates) destroys an electron with spin $\sigma$ on site $i$ and
$n_{i\sigma}$ is the corresponding number operator. $U$ is the on-site Coulomb repulsion. 
We consider hopping $t$ between nearest-neighbors $\langle ij \rangle$ and hopping $t'$ between
next-nearest-neighbors $\langle\langle il \rangle\rangle$. 
We show results for $t=1$, $t'=0.3$ and $U=8$, which are
realistic values for cuprates~\cite{eskes,maekawa,LDA}. 
We find PS for values of the filling  smaller than one,
which for positive  $t'$  corresponds to the electron doped cuprates.

The DCA is an extension of the Dynamical Mean Field Theory
(DMFT)\cite{DMFT}.  The DMFT maps the lattice problem to an impurity
embedded self-consistently in a host and therefore neglects spatial
correlations.  In the DCA we assume that correlations are short-ranged 
and map the original lattice model onto a periodic cluster of size 
$N_c=L_c\times L_c$ embedded in a self-consistent host.  Thus, correlations 
up to a range $\xi\lesssim L_c$ are treated accurately, while the physics 
on longer length-scales is described at the mean-field level.    We solve 
the cluster problem using quantum Monte Carlo (QMC) \cite{jarrell:dca3}.
The cluster self-energy is used to calculate the properties of the host,
and this procedure is repeated until a self-consistent convergent
solution is reached.

Unlike most of the other numerical calculations on PS, which study systems 
with a fixed number of particles, our calculations are done in the grand 
canonical ensemble and in the thermodynamic limit.
Therefore, unlike in finite cluster
calculations, we do not encounter any particular difficulty associated with the
small doping regime. Phase separation is explored by calculating  the filling  dependence on the 
chemical potential and  the   charge susceptibility (or compressibility),  
$\chi_{charge}=\frac{dn}{d \mu}$.

\paragraph*{Results}

\begin{figure}
\centering 
\includegraphics*[width=3.3in]{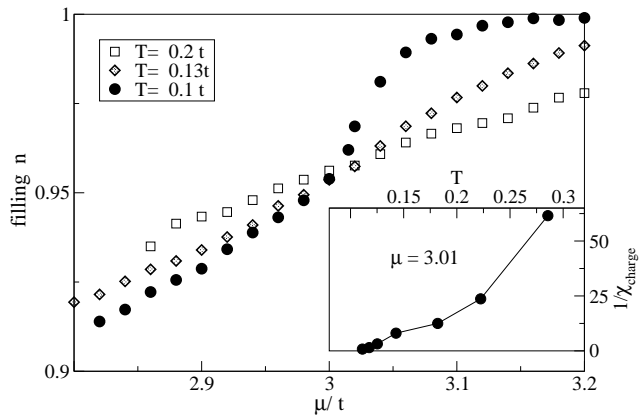} 
\caption{$N_c=8$ results. Filling $n$ versus chemical potential when $T>T_c\approx 0.1~t$.
Inset: Inverse of the charge susceptibility $\chi_{charge}$ versus temperature for fixed
chemical potential $\mu=\mu_c$.} 
\label{fig:hysg}
\end{figure}

First we consider  the  case of an eight site ($N_c=8$)  cluster.  The  
filling as a function of the  chemical potential is plotted in  
Fig.~\ref{fig:hysg} for different temperatures~\footnote{The statistical error 
bars on the densities are of the order of the symbol size or smaller.  In the
DCA, the error bars for the lattice susceptibilities can only be obtained 
by repeated runs. Due to the computational expense of this procedure, it 
was generally not done, except where specified.}. 
Note that that at small doping with lowering temperature the charge 
susceptibility is increasing and diverging at a critical point ($\delta_c,
\mu_c,T_c $). The  divergence of the charge susceptibility is illustrated in
the inset.  It is a clear 
indication that the  filling is unstable and the system is subject to phase 
separation into regions with different hole density. The critical point is 
characterized by the  temperature $T_c \approx 0.10~t$ and the doping  
$\delta_c \approx 4.5\%$.

\begin{figure}
\centering 
\includegraphics*[width=3.3in]{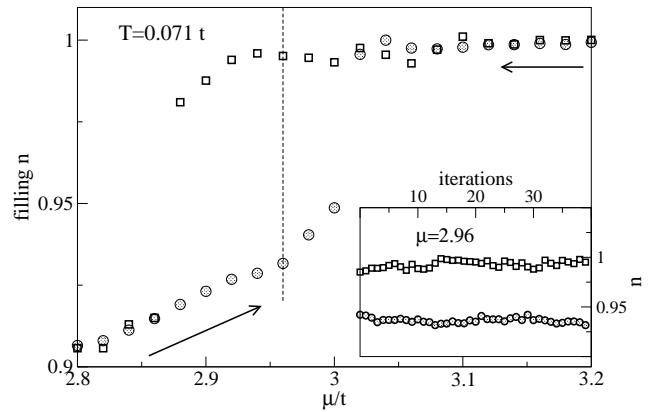} 
\caption{$N_c=8$ results. Filling $n$ versus chemical potential below 
T$_c$, at $T=0.071~t$.  Two solutions describing a hysteresis are found, 
one incompressible with $n \approx 1$ (squares) and a  doped one (circles).  
Inset: stability 
of the two solutions versus DCA iterations when $\mu=2.96 t$ (middle
of the hysteresis, corresponding to the dotted  line in the main figure). } 
\label{fig:hysl}
\end{figure}

For temperatures smaller than $T_c$ and for values of the chemical potential
close to $\mu_c$  the DCA calculation provides two distinct solutions for 
the same value of $\mu$. As mentioned before, the  DCA equations are solved
self-consistently starting with an initial guess for the self-energy, usually 
zero or that from a larger temperature or a perturbation theory result.  In 
most of the situations an unique solution is obtained independent of the 
starting guess. This is the case at doping values far from  $\delta_c$ such 
as $0\%$ (undoped) or $10\%$ doping.  However, close to  $\mu_c$  we find 
that the final solution is dependent on the starting point. If one uses as 
the initial input the self-energy corresponding to the undoped solution ($n=1$),  
then $n$ versus $\mu$ will look as the upper curve (squares) in 
Fig.~\ref{fig:hysl}.  On the other hand if the starting self-energy is the 
one corresponding to the large doped solution ($n<1$), $n$ versus $\mu$ will 
be described by the lower curve  (circles) in  Fig.~\ref{fig:hysl}. In both 
cases, the fully converged self energy of the previous point is used to 
initialize the calculation.  Thus, below $T_c$ the  filling as a 
function of the chemical potential displays a hysteresis.

Simple thermodynamic ideas may be used to interpret these results.  A  
hysteresis implies the existence of a metastable state and it is observed 
in many systems which suffer a first order transition, a common example 
being magnetization versus the applied magnetic field  ($M(H)$) in magnetic 
materials.  However in the real systems, after a sufficient time, the 
fluctuations  always drive the system to the stable solution (the equilibrium 
solution) and the hysteresis becomes a discontinuity characteristic to  
first order transitions.  In our case, due to the mean-field coupling of 
the cluster to the effective medium, the hysteresis is stable. This is shown 
in the inset of  Fig.~\ref{fig:hysl} where a large number of iterations in 
the  self-consistent process is considered.

\begin{figure}
\centering 
\includegraphics*[width=3.3in]{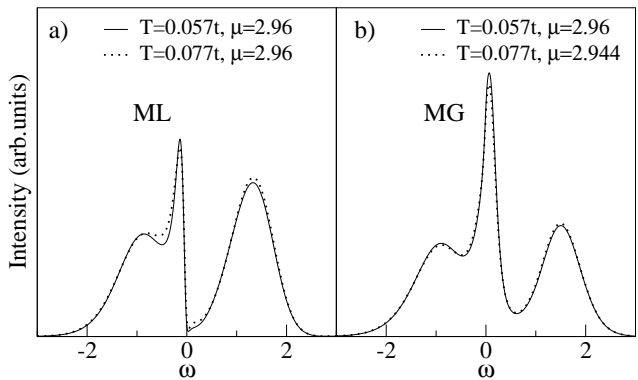} 
\caption{a) DOS for ML solution at $T=0.077t$ (dotted line)
and $T=0.057t$ (full line). b) DOS for MG solution at $T=0.077t$ (dotted line)
and $T=0.057t$ (full line).} 
\label{fig:DOS}
\end{figure}

By analogy with the liquid-gas system discussed below, 
we label the two states found  for $T<T_c$ as  Mott liquid (ML) and 
Mott gas (MG).  The Mott liquid is incompressible and insulating.  Both the 
compressibility and doping of the ML are small and decrease with decreasing
temperature.  Its density of states at the Fermi surface develops a gap 
with lowering temperature characteristic of an insulator, as seen in 
Fig.~\ref{fig:DOS}-a.   The MG is compressible and metallic.  The DOS is 
peaked at the chemical potential and increases with lowering temperature, 
see Fig.~\ref{fig:DOS}-b).   Consistent with the narrow peak width,  the 
MG  is a strongly correlated state with a small value of the double occupancy 
($\langle n_{i\uparrow} n_{i\downarrow} \rangle / n \approx 0.04$ at
$T=0.077 t$) and strong AF correlations.

\begin{figure}
\centering 
\includegraphics*[width=3.3in]{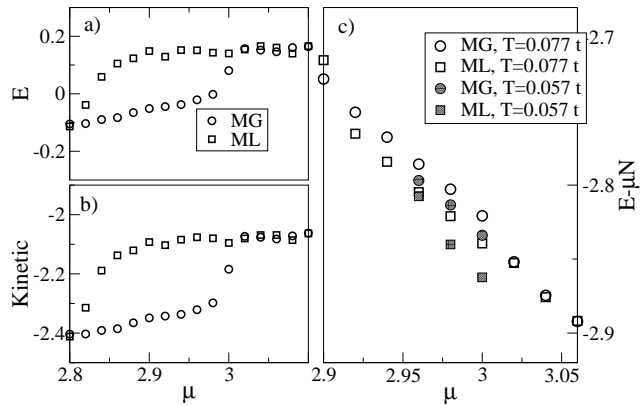} 
\caption{Energy per site versus $\mu$ for the
two solutions: a) total energy $E=\langle H \rangle$, (see Eq.~\ref{eq:HM}) at $T=0.077 t$, 
b) Kinetic energy $=\langle H_{kin} \rangle$ (see Eq.~\ref{eq:hams}) at $T=0.077 t$, 
c) $E-\mu N$  at $T=0.077 t$ and $T=0.057 t$.}
\label{fig:energies}
\end{figure}

The stable solution below $T_c$, ML or MG, is the one with lower free 
energy, $F=E-\mu N -TS$.  Unfortunately, due to the mean-field character 
of the DCA, the self-consistent solution is not necessarily the equilibrium 
state, and the QMC method does not allow the calculation of the entropy.
Therefore the determination of the critical $\mu$ where the jump in 
$N(\mu)$ should take place is difficult to identify.  However, the  
calculation of the energy provides valuable information about the transition 
mechanism. The energy plotted versus $\mu$ displays a cusp at $\mu_c$ when 
$T=T_c$ (not shown).  Below $T_c$, the energy is hysteretic. As can be seen 
in Fig.~\ref{fig:energies}-a at fixed $\mu$ the energy of the gas phase is 
much smaller, due to the large gain in kinetic energy (see 
Fig.~\ref{fig:energies}-b) produced by the next-nearest-neighbor hopping 
$t'$ as we will discuss.
On the other hand, the term $-\mu N$ will favor the ML state since it 
has a larger filling. In fact we find that the difference between 
$E-\mu N$ for the two solutions is small, with the ML state being 
favored for larger values of $\mu$. When the chemical potential is decreased 
the system will be driven to the MG state by both the lower kinetic energy
and the larger, presumably, entropy characteristic to MG state. Therefore,
for $T<T_c$, we expect the jump in $n$  will move to lower values of 
$\mu$ as the temperature is lowered.

\begin{figure}
\centering
\includegraphics*[width=3.3in,height=2.2in]{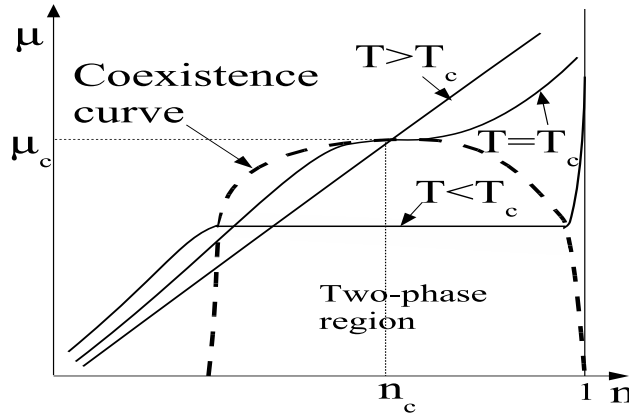} 
\caption{Schematic representation of the phase diagram.} 
\label{fig:cartoon}
\end{figure}

One can notice that a phase diagram with these characteristics bears a 
striking similarity to the phase diagram of a classical liquid gas 
mixture~\cite{he_stanley}, where $\mu$ plays the role of pressure.  A 
cartoon which summarizes our results and illustrates this similarity is 
shown in Fig~\ref{fig:cartoon}.  At high $T$, $n$ versus $\mu$ is linear, 
since correlations are irrelevant.  As the temperature is lowered, $n(\mu)$ 
becomes nonlinear due to correlation effects.  At $T_c$, $dn/d \mu$ diverges.  
Below $T_c$ the hysteresis appears.  Upon lowering the temperature the 
hysteresis broadens and the MG (ML) solution shifts to slightly larger 
(smaller) dopings.  As $T \rightarrow 0$, the entropy term becomes smaller 
and the chemical potential $\mu_c$ where the jump takes place in the real 
solution should move to smaller values. If a fixed $N$ is imposed when 
$T<T_c$ in the two-phase parameter regime, the system will separate into 
distinct ML and MG regions.

\begin{figure}[t]
\centering
\includegraphics*[width=3.3in]{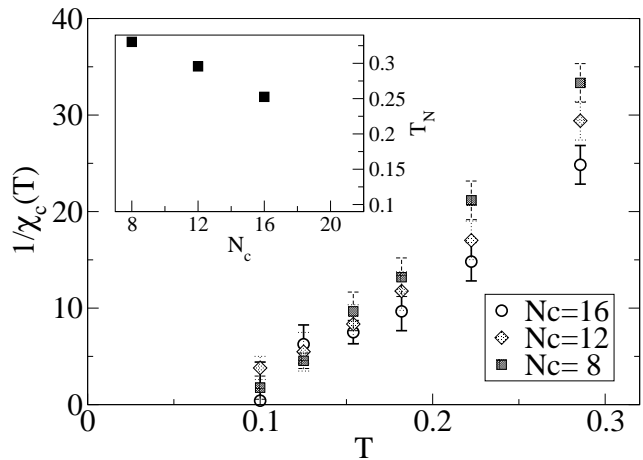}
\caption{Inverse charge susceptibility $1/\chi_{charge}$ versus temperature for  
$N_c=8$ (squares), $N_c=12$
(diamonds) and $N_c=16$ (circles) site clusters
at $5\%$ doping. Inset: The AF temperature $T_N$ versus cluster size $N_c$.} 
\label{fig:clusters}
\end{figure}

Our calculations assume a paramagnetic host which implies that the
range of possible AF order is restricted to the cluster size. For the
$N_c=8$ cluster we find PS below the AF critical temperature $T_N$,
the temperature where the AF spin susceptibility is diverging and the
AF correlations range reaches the cluster size. Therefore it is
important to address the role of AF correlations on phase
separation. For this we investigate the behavior of the critical
temperatures $T_c$ and $T_N$ when the cluster size increases. In the
inset of Fig.~\ref{fig:clusters} one can see that at $5\%$ doping
$T_N$ decreases rapidly with increasing cluster size.  On the other
hand, the $N_c=12$ and $N_c=16$ site clusters display a divergent
charge susceptibility roughly at the same $T_c$ as the $N_c=8$
cluster, $T_c \approx 0.1 t$, as shown in Fig.~\ref{fig:clusters}. The
rapid decrease of $T_N$ with $N_c$ and the fact that $T_c$ is nearly independent
on $N_c$  indicates that PS may persist in larger
clusters at a temperature higher than $T_N$ where the range of AF
correlations is smaller than the cluster size.  However, we must
mention that the calculations on $N_c=12$ and $N_c=16$ clusters, close
to the PS temperature, are extremely difficult. This region of
parameter space is characterized by very strong critical behavior
(presumably because a larger cluster implies a weaker hybridization
with the effective medium, i.e. the results are less ``mean-field''),
a severe minus sign problem, and extremely large auto correlation
times between measurements.  Consequently, the error bar in the
filling and the charge susceptibility is increasingly large for the
low temperature points in Fig.~\ref{fig:clusters} and it is difficult
to obtain converged solutions.  Therefore, besides a rough estimation
of $T_c\approx 0.1 t$ it is difficult to make other quantitative
estimates for the critical parameters. Low temperature calculations on
larger clusters inside the critical region where a hysteresis is
expected are not possible due to the severe sign problem which appears
in the QMC calculation.

We find PS only when the next-nearest-neighbor hopping $t'>0$ and the filling
$n<1$. A finite $t'$ in Hubbard and t-J models is known to give rise to a 
strong asymmetry between electron-doped ($t'>0$) and hole-doped ($t'<0$)
systems~\cite{maekawa,gooding,maekawa1,2band}.  In exact diagonalization studies 
on small clusters~\cite{riera,itoh,gagliano,maekawa,gooding} it was shown that,
due to the kinetic energy gain, the motion of holes caused by a positive $t'$ 
stabilizes antiferromagnetic configurations~\cite{maekawa,gooding}.
Even though exact diagonalization on systems with four holes suggests that $t'$ 
is not favorable to hole clustering~\cite{itoh,gooding} the tendency to PS was 
noted in Ref.~\cite{gooding}.  Our calculations also do not indicate hole clustering 
but rather formation of a $8\%-10\%$ doped state with strong AF correlations and 
low-kinetic energy.  Presumably for this value of the doping the effect of $t'$ on 
the kinetic energy is the most significant.

For smaller values of $t'$, PS takes place at lower temperatures.  For instance 
when  $t'=0.1 t$, the system shows PS at $T_c=0.055 t$ for the $N_c=8$ cluster.  
For $U<W$ we found no sign of PS for temperatures above $0.04~t$.  The charge 
susceptibility behavior suggests that PS is not favored when $t'<0$ and $n<1$, 
in agreement with  exact diagonalization results~\cite{gooding,maekawa2} which 
show that in this case the effect of $t'$ is to push the holes apart from each 
other.

Our results imply phase separation into two regions with different 
electronic density. However, even without considering long range order, 
we cannot exclude the  possibility that PS competes with the 
formation of different charge  patterns such as  stripes or checkerboard.
The investigation of these instabilities would require calculations on much 
larger clusters, able to commensurate these patterns, 
which are unfeasible at the moment. 

It would be interesting to investigate the competition between PS and
d-wave superconductivity in the Hubbard model. However this implies
a region of the parameter space not accessible to our method.
Calculations on clusters larger than $N_c=8$ show PS
but the sign problem precludes access to
temperatures where the superconductivity is expected.
On the other hand, calculations on the small $2 \times 2$ cluster,
where the sign problem is mild, show d-wave superconductivity for finite $t'$
but no definite evidence for PS,  even though  the charge susceptibility is strongly  
increased when a positive $t'$ is considered~\cite{2band}.

\paragraph*{Conclusions}

With the DCA we show that the Hubbard model with a positive
next-nearest-neighbor hopping displays PS for values of the filling
slightly smaller than one. Our results suggest that the PS is driven
by the desire to form slightly doped ($\approx 8\%-10\%$) regions with
low-kinetic energy and strong antiferromagnetic correlations.  The phase
diagram is similar to that of the liquid-gas mixture, showing a second
order critical point and a first order transition from a Mott gas to a
Mott liquid state below $T_c$.

\paragraph*{Acknowledgment}

We acknowledge useful discussions with S.R. White.  This research was 
supported by NSF Grants DMR-0312680 and DMR-0113574, by CMSN grant DOE DE-FG02-04ER46129 
and was supported in part by NSF cooperative agreement SCI-9619020 through resources provided by 
the San Diego Supercomputer Center. Part of the computation was also performed at the
Center for Computational Sciences at the Oak Ridge National Laboratory.


\begin{thebibliography}{99}
\expandafter\ifx\csname natexlab\endcsname\relax\def\natexlab#1{#1}\fi
\expandafter\ifx\csname bibnamefont\endcsname\relax
  \def\bibnamefont#1{#1}\fi
\expandafter\ifx\csname bibfnamefont\endcsname\relax
  \def\bibfnamefont#1{#1}\fi
\expandafter\ifx\csname citenamefont\endcsname\relax
  \def\citenamefont#1{#1}\fi
\expandafter\ifx\csname url\endcsname\relax
  \def\url#1{\texttt{#1}}\fi
\expandafter\ifx\csname urlprefix\endcsname\relax\def\urlprefix{URL }\fi
\providecommand{\bibinfo}[2]{#2}
\providecommand{\eprint}[2][]{\url{#2}}

\bibitem{stripe} J. M. Tranquada, B. J. Sternlieb, J. D. Axe, Y. Nakamura, S. Uchida, 
Nature (London) {\bf 375}, 561 (1995); J. M. Tranquada, J. D. Axe, N. Ichikawa, Y. Nakamura, S. Uchida
and B. Nachumi, Phys. Rev. B {\bf 54}, 7489 (1996); 
J. M. Tranquada, J. D. Axe, N. Ichikawa, A. R. Moodenbaugh, Y. Nakamura, and S. Uchida,
Phys. Rev. Lett.{\bf 78}, 338 (1997).
\bibitem{stm} J. E. Hoffman, E. W. Hudson, K. M. Lang, V. Madhavan, H. Eisaki, 
S. Uchida, and J. C. Davis, Science {\bf 295}, 466 (2002); 
C. Howald, H. Eisaki, N. Kaneko, M. Greven, and A. Kapitulnik,
Phys. Rev. B {\bf 67}, 014533 (2003);
Michael Vershinin, Shashank Misra, S. Ono, Y. Abe, Yoichi Ando, and Ali Yazdani,
Science {\bf 303}, 1995 (2004);
T. Hanaguri, C. Lupien, Y. Kohsaka, D.-H. Lee, M. Azuma, M. Takano, H. Takagi, J. C. Davis,
Nature (London) {\bf 430}, 1001 (2004);
K. McElroy, D. H. Lee, J. E. Hoffman, K. M. Lang, J. Lee, E. W. Hudson, H. Eisaki, S. Uchida, 
and J. C. Davis, Phys. Rev. Lett. {\bf 94}, 197005 (2005).
\bibitem{PStoCDW}  S.A. Kivelson and V.J. Emery, in {\em Strongly Correlated Electronic Materials: The Los Alamos Symposium 1993}, edited by K. S. Bedell, Z. Wang, and D. E. Meltzer (Addison-Wesley, Redwood City, 1994); 
G. Seibold, C. Castellani, C. Di Castro, and M. Grilli, Phys. Rev. B {\bf 58}, 13506 (1998).
\bibitem{EKL} V. J. Emery, S. A. Kivelson and H. Q. Lin, Phys. Rev. Lett. {\bf 64}, 475, (1990).
\bibitem{HellbergManousakis} C. S. Hellberg and E. Manousakis, Phys. Rev. Lett. {\bf 78}, 4609, (1997); Jung Hoon Han, Qiang-Hua Wang, and Dung-Hai Lee, 
Int. J. Mod. Phys. B {\bf 15}, 1117 (2001).
\bibitem{u1sboson} %T.-H. Gimm and S.-H. S. Salk, 
Tae-Hyoung Gimm and Sung-Ho SuckSalk, Phys. Rev. B {\bf 62}, 13930, (2000).
\bibitem{putikka} W. O. Putikka and M. U. Luchini,  Phys. Rev. B {\bf 62}, 1684, (2000).
\bibitem{shih} C. T. Shih, Y. C. Chen and T. K. Lee, Phys. Rev. B {\bf 57}, 627, (1998).
\bibitem{moreo} A. Moreo, D. Scalapino, and E. Dagotto, Phys. Rev. B {\bf 43}, 11442, (1991).
\bibitem{sorella} Federico Becca, Massimo Capone and Sandro Sorella, Phys. Rev. B {\bf 62}, 12700, (2000).
\bibitem{su} Gang Su, Phys. Rev. B  {\bf 54}, R8281, (1996).
\bibitem{gehlhoff} Lew Gehlhoff, J. Phys: Condensed Matter  {\bf 8}, 2851 (1996).
\bibitem{pruschke} R. Zitzler, Th. Pruschke, and R. Bulla, 
Eur. Phys. J. B {\bf 27}, 473, (2002).
\bibitem{arrigoni} M. Aichhorn and E. Arrigoni, Europhys. Lett. {\bf 71}, 117 (2005);
M, Aichhorn, E. Arrigoni, M. Potthoff and  W. Hanke, preprint, cond-mat/0511460  (2005).
\bibitem{hettler:dca} M. H. Hettler, A. N. Tahvildar-Zadeh, M. Jarrell, T. Pruschke, and H. R. Krishnamurthy,  Phys. Rev. B {\bf{58}}, R7475 (1998);
M. H. Hettler, M. Mukherjee, M. Jarrell, and H. R. Krishnamurthy, 
Phys.\ Rev.\ B {\bf{61}}, 12739 (2000);  
T. Maier, M. Jarrell, T. Pruschke, and J. Keller,  Eur. Phys. J. B {\bf 13}, 613 (2000).
\bibitem{maier:rev} T. Maier, M. Jarrell, T. Pruschke, and M. H. Hettler,  
Rev. Mod. Phys. {\bf{77}}, 1027 (2005).
\bibitem{eskes} H. Eskes, G. A. Sawatzky, and L. Feiner, Physica C, {\bf 160}, 424 (1989).
\bibitem{maekawa} T. Tohyama and S. Maekawa, J. Phys. Soc. Japan, {\bf 59}, 1760, (1990);
T. Tohyama and S. Maekawa, Phys. Rev. B {\bf 49}, 3596 (1993).
\bibitem{LDA} Mark S. Hybertsen, E. B. Stechel, M. Schluter, and D. R. Jennison,
Phys. Rev. B {\bf 41}, 11068, (1990).
\bibitem{DMFT} Th. Pruschke, M. Jarrell, and J.K. Freericks, Advances in Physics {\bf 44}, 187 (1995);
 A. Georges G.Kotliar, W.Krauth and M.Rozenberg, Rev.Mod.Phys. {\bf 68}, 13 (1996).
\bibitem{jarrell:dca3} M. Jarrell, Th. Maier, C. Huscroft, and S. Moukouri, 
Phys.Rev. {\bf B 64}, 195130 (2001).
\bibitem{gooding} R. J. Gooding, K. J. E. Vos and P. W. Leung, 
Phys. Rev. B {\bf 50}, 12866, (1994).
\bibitem{maekawa1}  T. Tohyama and S. Maekawa, Supercond. Sci. Technol.  {\bf 13}, R17, (2000)
\bibitem{2band}  A. Macridin, M. Jarrell, Th. Maier, and G. A. Sawatzky,  Phys. Rev. B {\bf 71}, 134527, (2005).
\bibitem{riera} J. A. Riera,  Phys. Rev. B {\bf 40}, R833, (1989)
\bibitem{itoh} Toshihiro Itoh, Masao Arai, and Takeo Fujiwara, 
Phys. Rev. B {\bf 42}, R4834 (1990).
\bibitem{gagliano} Eduardo Gagliano, Silvia Bacci, and Elbio Dagotto,  Phys. Rev. B {\bf 42}, 6222, (1990).
\bibitem{he_stanley} H.E.\ Stanley, {\em Introduction to Phase Transitions and 
Critical Phenomena}, (Oxford Univ. Press, New York, 1971).
\bibitem{maekawa2} T. Tohyama and S. Maekawa, Phys. Rev. B {\bf 67}, 092509 (2003).
\end{thebibliography}
\end{document}